\documentclass[twocolumn,tighten,twocolumn]{aastex701}
\usepackage{wrapfig}
\usepackage{graphics,graphicx}
\usepackage{epstopdf}
\usepackage[utf8]{inputenc}
\usepackage{tikz}
\usetikzlibrary{shapes,arrows}
\usepackage{amssymb, amsmath, amsthm}
\usepackage[normalem]{ulem}
\usepackage{float}
\usepackage{color}
\usepackage{xcolor}
\usepackage{amsmath} 
\usepackage{graphicx}
\usepackage{booktabs}
\usepackage{multirow}
\usepackage{soul}
\definecolor{ultramarine}{rgb}{0.01, 0.64, 0.86} 
\def\green#1 {{\textcolor{ultramarine}{#1}}\ }

\begin{document}
\defcitealias{2026ApJS..282...39J}{Jiao26}

\title{Low Star Formation Efficiency in M31: Evidence for a Deficit of Bound Gas}

\correspondingauthor{Sihan Jiao, Hauyu Baobab Liu}
\email{sihanjiao@nao.cas.cn, baobabyoo@gmail.com}

\author{Fangyuan Deng}
\email{dengfangyuan21@mails.ucas.ac.cn}
\affiliation{University of Chinese Academy of Sciences, Beijing 100049, China}
\affiliation{
National Astronomical Observatories, Chinese Academy of Sciences, 20A Datun Road, Chaoyang District, Beijing 100012, China
}
\email{dengfangyuan21@mails.ucas.ac.cn}

\author[0000-0002-9151-1388]{Sihan Jiao}
\affiliation{
National Astronomical Observatories, Chinese Academy of Sciences, 20A Datun Road, Chaoyang District, Beijing 100012, China
}
\affiliation{
Max Planck Institute for Astronomy, Konigstuhl 17, D-69117 Heidelberg, Germany
}
\email{sihanjiao@nao.cas.cn}

\author[0000-0003-2300-2626]{Hauyu Baobab Liu}
\affiliation{Department of Physics, National Sun Yat-Sen University, No. 70, Lien-Hai Road, Kaohsiung City 80424, Taiwan, R.O.C.}
\affiliation{Center of Astronomy and Gravitation, National Taiwan Normal University, Taipei 116, Taiwan}
\email{baobabyoo@gmail.com}

\author{Jingwen Wu}
\affiliation{University of Chinese Academy of Sciences, Beijing 100049, China}
\affiliation{
National Astronomical Observatories, Chinese Academy of Sciences, 20A Datun Road, Chaoyang District, Beijing 100012, China
}
\email{jingwen@nao.cas.cn}

\author{Henrik Beuther}
\affiliation{
Max Planck Institute for Astronomy, Konigstuhl 17, D-69117 Heidelberg, Germany
}
\email{beuther@mpia.de}

\author{Fanyi Meng}
\affiliation{New Cornerstone Science Laboratory, Department of Astronomy, Tsinghua University, Beijing 100084, China}
\affiliation{University of Chinese Academy of Sciences, Beijing 100049, China}
\email{mengfanyi@mail.tsinghua.edu.cn}

\author{Caroline Gieser}
\affiliation{
Max Planck Institute for Astronomy, Konigstuhl 17, D-69117 Heidelberg, Germany
}
\email{gieser@mpia.de}

\author{Yuxin Lin}
\affiliation{Max-Planck-Institut f\"ur Extraterrestrische Physik, Giessenbachstr. 1, D-85748 Garching bei M\"unchen, Germany}
\email{ylin@mpe.mpg.de}

\author{Di Li}
\affiliation{New Cornerstone Science Laboratory, Department of Astronomy, Tsinghua University, Beijing 100084, China}
\affiliation{
National Astronomical Observatories, Chinese Academy of Sciences, 20A Datun Road, Chaoyang District, Beijing 100012, China
}
\email{dili@mail.tsinghua.edu.cn}

\author[0000-0002-9390-9672]{Chao-Wei Tsai} 
\affiliation{
National Astronomical Observatories, Chinese Academy of Sciences, 20A Datun Road, Chaoyang District, Beijing 100012, China
} 
\affiliation{Institute for Frontiers in Astronomy and Astrophysics, Beijing Normal University, Beijing 102206, China} 
\affiliation{School of Astronomy and Space Science, University of Chinese Academy of Sciences, Beijing 100049, 
China}
\email{cwtsai@nao.cas.cn}

\author{Junzhi Wang}
\affiliation{
School of Physical Science and Technology, Guangxi University, Nanning 530004, China
}
\email{junzhiwang@gxu.edu.cn}

\author{Zhi-Yu Zhang}
\affiliation{School of Astronomy and Space Science, Nanjing University, Nanjing 210093, China}
\affiliation{Key Laboratory of Modern Astronomy and Astrophysics, Ministry of Education, Nanjing 210093, China}
\email{zzhang@nju.edu.cn}

\author[0000-0001-5950-1932]{Fengwei Xu}
\affiliation{
Max Planck Institute for Astronomy, Konigstuhl 17, D-69117 Heidelberg, Germany
}
\email{fengwei@mpia.de}

\author[0000-0002-8760-6157]{Jakob den Brok}
\affiliation{
Max Planck Institute for Astronomy, Konigstuhl 17, D-69117 Heidelberg, Germany
}
\email{jakob.denbrok@gmail.com}

\author{Linjing Feng}
\affiliation{
National Astronomical Observatories, Chinese Academy of Sciences, 20A Datun Road, Chaoyang District, Beijing 100012, China
}
\affiliation{University of Chinese Academy of Sciences, Beijing 100049, China}
\email{ljfeng@nao.cas.cn}

\author{Zongnan Li}
\affiliation{Korea Astronomy and Space Science Institute, 776 Daedeok-daero, Yuseong-gu, Daejeon 34055, Republic of Korea}
\affiliation{National Astronomical Observatory of Japan, 2-21-1 Osawa, Mitaka, Tokyo, 181-8588, Japan}
\email{zongnanli42@gmail.com}

\author{Yuxiang Liu}
\affiliation{University of Chinese Academy of Sciences, Beijing 100049, China}
\affiliation{
National Astronomical Observatories, Chinese Academy of Sciences, 20A Datun Road, Chaoyang District, Beijing 100012, China
}
\email{liuyuxiang22@mails.ucas.ac.cn}

\begin{abstract}

The Andromeda galaxy (M31) exhibits a systematically low star formation efficiency (SFE) compared to other star-forming galaxies. 
Star formation is thought to be regulated by the mass of gravitationally bound gas.
To investigate whether the low SFE in M31 originates from a scarcity of such bound gas, we analyze the dynamical state of its giant molecular cloud (GMC) population by estimating their virial parameters ($\alpha_{\text{vir}}$).
Using a JCMT-SCUBA2 850\,\textmu m dust-continuum GMC catalog as the parent sample, we combine it with IRAM 30m CO(1-0) data to define a working sample of 173 GMCs and derive their kinematic properties.
Most analyzed GMCs exhibit high $\alpha_{\text{vir}}$, indicating that they are predominantly gravitationally unbound at the $\sim86$-pc resolution of our observations.
While no statistically significant global correlation is found between SFE and $\alpha_{\text{vir}}$ for the full cloud sample, the upper quantiles of the SFE distribution show progressively stronger negative trends with increasing $\alpha_{\text{vir}}$.
This upper-tail trend may arise from the combined effects of the virial parameter limiting the upper range of SFE values accessible to the GMC population and feedback-driven disruption reducing the number of high-SFE, high-$\alpha_{\rm vir}$ CO-bright clouds in the observed sample.
The substantial scatter in the SFE-$\alpha_{\text{vir}}$ plane likely reflects the diverse evolutionary stages of GMCs, the temporal mismatch between gas and star formation tracers, and the limitations of a cloud-averaged virial parameter as a tracer of internal dynamical structure, making a simple one-to-one correspondence between $\alpha_{\text{vir}}$ and instantaneous SFE difficult to recover for individual GMCs.

\end{abstract}

\keywords{galaxies: individual (M31) ---  galaxies: ISM --- galaxies: structure --- ISM: clouds}  

\section{Introduction}\label{section:introduction}

Giant Molecular Clouds (GMCs) are the primary sites of star formation in galaxies and therefore play an important role in regulating the buildup of stellar mass and the evolution of the interstellar medium (ISM) \citep[e.g.,][]{McKee2007ARA&A,Kennicutt2012}. 
The conversion of gas into stars within GMCs is regulated by a variety of physical processes that are still being actively studied \citep[e.g.,][]{2023ASPC..534....1C}.
While studies within the Milky Way offer high spatial resolution, they are often limited by distance uncertainties and line-of-sight confusion due to the edge-on view of the Galactic disk. 
External galaxies provide a more complementary perspective, allowing molecular cloud populations to be studied within a well-defined galactic environment.
Among them, the Andromeda galaxy (M31), owing to its proximity \citep[$d$ $\sim$\,766\,kpc,][]{Lee2023ApJ} and moderate inclination, provides an ideal laboratory for studying GMC properties and their connection to star formation on a global scale with reduced projection effects compared to our Galaxy.

Extensive multi-wavelength surveys have characterized the ISM and stellar populations of M31. 
These include observations from {\it GALEX} \citep{ford2013ApJ}, {\it Spitzer} \citep{Barmby2006ApJ,Gordon2006ApJ}, {\it Herschel} \citep{Fritz2012A&A,Draine2014ApJ}, {\it Planck} \citep{Planck2015A&A}, and VLA \citep{Koch2021MNRAS,Koch2025arXiv}, allowing for detailed studies of dust and gas properties and star-formation activities. 
Recently, \citeauthor{2026ApJS..282...39J} (\citeyear{2026ApJS..282...39J}, hereafter \citetalias{2026ApJS..282...39J}) presented a comprehensive catalog of GMCs in M31 identified via 850 $\mu$m continuum emission using the JCMT SCUBA-2. 
By examining the relationship between star formation rate surface density ($\Sigma_{\text{SFR}}$) and gas surface density ($\Sigma_{\text{gas}}$), they found that the star formation efficiency (SFE) of M31 is systematically lower than that in other nearby star-forming galaxies. 

The globally low SFE of M31 has also been reported in previous studies \citep[e.g.,][]{Leroy2008AJ, Sun2023ApJ...945L..19S}. 
This relatively low star formation rate places M31 in the Green Valley of the color-magnitude diagram, representing a transitional stage between actively star-forming blue galaxies and quiescent red systems \citep[e.g.,][]{2011ApJ...736...84M}.
In the context of galaxy quenching, two broad mechanisms are commonly invoked: the removal of cold gas or the suppression of its collapse \citep[e.g.,][]{2009ApJ...707..250M,2015Natur.521..192P,2018NatAs...2..695M}.
For M31, however, gas depletion is unlikely to be the dominant factor, as the galaxy still contains a substantial reservoir of cold molecular gas \citep[e.g.,][]{Draine2014ApJ, Smith2012ApJ, 2025A&A...701A.152J}. 
Instead, the low SFE might suggest that the main limitation likely lies in the ability of the cold gas to become gravity dominant and subsequently collapse to form stars.

To investigate the origin of this inefficiency, it is necessary to move from galaxy scales to the scale of individual GMCs. 
On these scales, star formation depends on the ability of gas to become gravitationally bound and collapse, while turbulence and magnetic fields regulate the balance between support and self-gravity \citep[e.g.,][]{McKee2007ARA&A,2005ApJ...630..250K,2014prpl.conf...77P}.
\citet{2021ApJ...920..126E} performed a comprehensive virial analysis on catalogs of molecular cloud structures in the Milky Way, finding that only a small fraction, approximately 19\%, of the molecular gas mass resides in gravitationally bound entities.
This result offers a substantial solution to the fundamental problem of slow star formation in our Milky Way.
\citet{2025A&A...701A.152J} used column density probability distribution functions (N-PDFs) to isolate the gravity dominant component of molecular clouds and showed that the amount of gravity dominant gas largely determines the star formation rate.
This relation has recently been found to persist in the extreme environment of the Galactic Central Molecular Zone, implying that environmental effects on star formation may primarily act by regulating the formation of self-gravitating gas \citep{Feng2026CMZ}.
These results suggest a plausible explanation for the low SFE in M31: 
although the galaxy contains a substantial molecular gas reservoir, only a limited fraction of this gas may become gravitationally bound, thereby restricting the material available for star formation.

Testing this hypothesis requires quantifying the dynamical state of GMCs in M31.
While N-PDF analysis can probe the bound gas fraction of the targeted clouds using either dust emission \citep[e.g.,][]{2025A&A...701A.152J} or combined CO isotopologue data \citep{Feng2026CO}, its application requires resolving internal cloud structures on scales smaller than $\sim1$\,pc, which remains challenging for current M31 observations. 
A widely used alternative diagnostic for assessing the dynamical state is the virial parameter, $\alpha_{\text{vir}} \equiv 2K / U_g$, where K and $U_g$ denote the kinetic energy and self-gravitational potential energy \citep{1992ApJ...395..140B}. 
This parameter characterizes the balance between internal kinetic motions and self-gravity and is commonly used as an indicator of the dynamical state of molecular gas.
Recent studies suggest that many molecular clouds in M31 exhibit virial parameters well above 2, indicating that they may be dynamically unbound \citep[e.g.,][]{2024ApJ...966..193L,2025MNRAS.538.2445D}. 
However, whether such a dynamical state is a pervasive property of GMCs across the entire galaxy, and whether it directly explains the globally low SFE of M31, remains to be systematically investigated.

In this work, we utilize the GMC sample from \citetalias{2026ApJS..282...39J} combined with molecular line data to derive their virial parameters. 
By analyzing $\alpha_{\text{vir}}$ and its correlation with SFE, we aim to investigate whether the internal dynamical state of the clouds regulates the efficiency of star formation in M31.
In Section \ref{section:data}, we describe the data and our GMC sample selection criteria.
We then display our findings in Section \ref{section:results} and discuss their implications in Section \ref{section:discuss}.

\section{Data and Catalog}\label{section:data}

\begin{figure}
    \includegraphics[width=\linewidth]{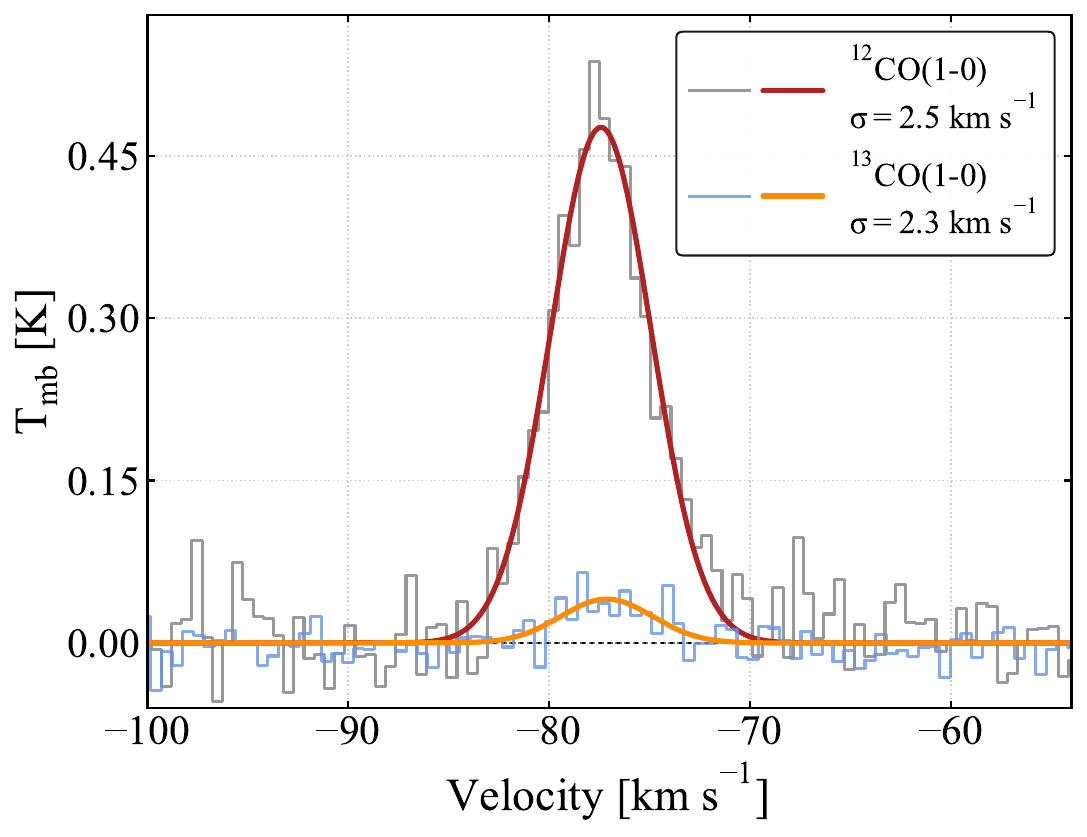}
    \caption{Spectra of the $^{12}$CO(1-0) and $^{13}$CO(1-0) emission lines toward the representative GMC J566, located at R.A.\,00:45:03.78 and Dec.\,+41:53:55.0.
    The data were taken from deep pointed IRAM 30\,m observations (Project ID: 059-18, PI: S. Jiao).
    The observed spectra are shown as stepped lines, while the curves represent the Gaussian fit profiles.
    The velocity dispersions derived from the fits are $\sigma_v = 2.50 \pm 0.07$\,km\,s$^{-1}$ and $\sigma_v = 2.29 \pm 0.40$\,km\,s$^{-1}$ for $^{12}$CO(1-0) and $^{13}$CO(1-0), respectively.
    }
    \label{fig:co_line_comparison}
\end{figure}

\begin{figure*}
    \centering
        \includegraphics[width=\linewidth]{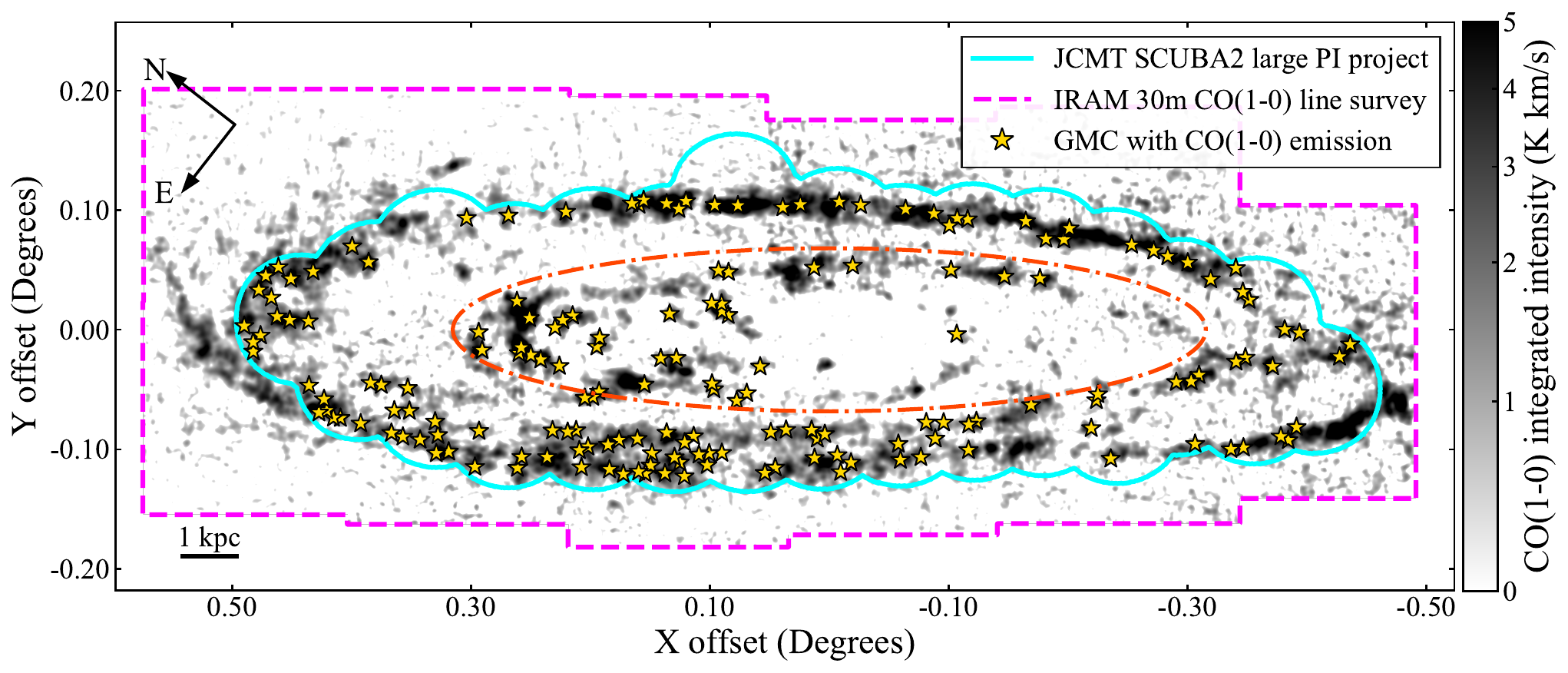}
        \caption{Spatial distribution of the 173 selected CO(1-0) tracing GMCs (gold stars) overlaid on the CO(1-0) integrated intensity image of M31. 
        The cyan contours outline the Field of View (FoV) of the JCMT SCUBA-2 survey, and the magenta dashed polygon indicates the FoV covered by the IRAM 30\,m CO(1-0) line survey. 
        The central orangered dashed ellipse marks the galactocentric radius $R_{\rm break} = 7$\,kpc (deprojected assuming an inclination of $77.5^\circ$), which indicates the transition radius of the virial parameter radial distribution observed in this work (see Section~\ref{subsubsec:vir_spa} and Figure~\ref{fig:vir_galacR}). 
        The black bar in the lower-left corner represents a physical scale of 1\,kpc.
        }
    \label{fig:co-source}
\end{figure*}

\subsection{Molecular Lines} \label{subsec:codata}

Measurements of gas kinematics, particularly the velocity dispersion within the GMCs, require molecular line observations. 
In this work, we utilize the CO(1-0) dataset from the IRAM 30\,m survey presented by \citet{Nieten2006A&A}, which covered the M31 disk.
This survey fully sampled an area of 2$^{\circ}$ $\times$ 0.5$^{\circ}$ toward M31 with an angular resolution of 23$^{\prime\prime}$ and a velocity channel width of 2.6\,km\,s$^{-1}$. The typical root-mean-square (RMS) noise is 25-33\,mK per 1\,MHz channel on the $T_{\rm mb}$ scale.
This CO(1-0) map provides complete coverage over the entire region of the JCMT-SCUBA2 dust map (see Section \ref{subsec:selectGMC}).

To validate the use of the $^{12}$CO(1-0) velocity dispersion as a reliable tracer of the internal kinematics of the GMCs, we utilize deep pointed observations from the IRAM 30\,m telescope (Project ID: 059-18, PI: S. Jiao).
These observations include simultaneous measurements of the $^{12}$CO(1-0), $^{13}$CO(1-0), and $^{12}$CO(2-1) lines toward selected pointings in M31, providing high-spectral-resolution spectra suitable for direct line-width comparisons.
Figure~\ref{fig:co_line_comparison} presents the spectra toward the representative GMC J566, located at R.A. 00:45:03.78 and Dec. +41:53:55.0.
The velocity dispersions derived from Gaussian fits to the $^{12}$CO(1-0) and $^{13}$CO(1-0) lines are comparable, suggesting that the $^{12}$CO(1-0) line widths are not significantly affected by optical depth broadening and can be used to reflect the bulk kinematic properties of the clouds.

\subsection{Parent Catalog and Sample Selection} \label{subsec:selectGMC}

Our analysis is built upon a parent sample of 572 GMCs cataloged by \citetalias{2026ApJS..282...39J}. 
This comprehensive catalog was constructed based on deep 850\,\textmu m dust continuum observations obtained by the JCMT Large Program (project code S19BP002, PI: Jingwen Wu), supplemented with \textit{Planck} 353\,GHz imaging. 
The JCMT survey maps a substantial portion of the M31 disk ($\sim$\,1.7$^{\circ}$ $\times$ 0.6$^{\circ}$) out to a galactocentric radius of $\sim$\,15\,kpc, as outlined by the cyan contour in Figure~\ref{fig:co-source}. It achieves a high sensitivity of $<$\,1.5\,mJy\,beam$^{-1}$ at 850\,\textmu m with a spatial resolution of $\sim$\,14$^{\prime\prime}$ (corresponding to $\sim$\,50\,pc). 
Using the \textit{Dendrogram} algorithm \citep{2008ApJ...679.1338R} on this dust map, \citetalias{2026ApJS..282...39J} successfully identified the boundaries and locations of the 572 GMCs.

For our subsequent analysis, we directly adopt the fundamental cloud properties derived in \citetalias{2026ApJS..282...39J} (see Table 2), specifically the effective radii ($R_{\rm pc}$) and the cloud masses ($M$).
Furthermore, to quantify the star formation activity associated with each GMC, we also adopt the star formation rate surface density ($\Sigma_{\text{SFR}}$) tabulated in the \citetalias{2026ApJS..282...39J} catalog. 
These $\Sigma_{\text{SFR}}$ values were originally extracted from the spatially resolved ($\sim$\,6$^{\prime\prime}$ resolution) $\Sigma_{\text{SFR}}$ map constructed by \citet{ford2013ApJ}. This map calibrates the total SFR using a combination of \textit{GALEX} Far-UV and \textit{Spitzer} MIPS 24\,\textmu m emissions \citep{Leroy2008AJ}.

While the \citetalias{2026ApJS..282...39J} catalog provides a comprehensive census of dust-identified clouds and their basic properties, calculating their dynamical states requires reliable kinematic measurements from spectral lines. 
Therefore, we performed a cross-match between the 572 GMCs and the molecular line data described in Section \ref{subsec:codata}. 
For each GMC, we used its cataloged central position and size to define the corresponding cloud aperture in the CO(1-0) data cube, and extracted a cloud-averaged CO spectrum by averaging over this aperture.
The observed velocity dispersion was derived from a Gaussian fit to the cloud-averaged spectrum and was then corrected for broadening caused by the spectral channel width following \citet{2006PASP..118..590R}.
The final working sample was then selected based on the following criteria:
\begin{enumerate}
    \item The extracted CO(1-0) spectrum must have a peak signal-to-noise ratio (S/N) greater than 3 and be well described by a single dominant velocity component. Spectra with insufficient S/N or clear multiple velocity peaks were excluded because a single-Gaussian fit would either be poorly constrained or would mix distinct velocity components into an artificially broad linewidth. This step reduces the sample from 572 to 243 GMCs.

    \item The observed velocity dispersion derived from the Gaussian fit, before correcting for spectral-channel broadening, must satisfy $\sigma_{v,{\rm obs}}>3\,{\rm km\,s^{-1}}$. This cut ensures that the CO line is sufficiently resolved relative to the 2.6\,km\,s$^{-1}$ channel width of the IRAM 30\,m CO(1-0) data. This criterion removes 8 additional GMCs.

    \item The angular size of the GMC must be larger than the CO(1-0) beam size ($\sim23^{\prime\prime}$, corresponding to a physical resolution of $\sim86$\,pc at the distance of M31). This criterion ensures that the CO linewidth is measured on a spatial scale comparable to that of the dust-defined cloud radius and mass used in the virial analysis. For sources smaller than the CO beam, the extracted spectrum would represent beam-averaged kinematics and could be affected by unresolved sub-beam structure, beam smearing, or low beam-filling factors, making the resulting linewidth difficult to interpret as a cloud-scale internal velocity dispersion. This criterion removes 59 additional GMCs.
\end{enumerate}
After these kinematic and resolution cuts, 176 GMCs remain. Three of them lack valid SFR measurements in the \citetalias{2026ApJS..282...39J} catalog and are therefore excluded from the SFE analysis. The final working sample thus consists of 173 GMCs with prominent, single-component CO(1-0) emission and the required measurements of cloud size, mass, linewidth, and SFR. 
Figure~\ref{fig:co-source} illustrates the spatial distribution of these selected sources overlaid on the velocity-integrated CO(1-0) intensity map presented by \citet{Nieten2006A&A}.

\section{Methods and Results}\label{section:results}

Recent studies suggest that the star formation rate is closely linked to the mass of gravity dominant gas \citep{2020ApJ...894..103E, 2021ApJ...920..126E, 2025A&A...701A.152J, Feng2026CMZ}. 
Motivated by this idea, we investigate whether the low SFE in M31 arises from a high degree of internal turbulence, which would limit the fraction of gas available for gravitational collapse.
To test this hypothesis, we calculate the virial parameter, $\alpha_{\text{vir}}$, and SFE for each GMC in our sample. 
The $\alpha_{\text{vir}}$ serves as a proxy for the dynamical state of the gas, allowing us to examine whether less gravitationally bound cloud tend to exhibit lower SFE. 

The fundamental cloud properties required for the analysis, namely the equivalent radii ($R_{\text{c}}$) and GMC masses ($M_{\text{c}}$), are directly adopted from the \citetalias{2026ApJS..282...39J} catalog.
The radius is defined based on the projected physical area ($A$) of the cloud identified by the \textit{Dendrogram} algorithm, expressed as $R_{\text{c}} = \sqrt{A/\pi}$, and this cloud radius is further corrected for the finite angular resolution through beam deconvolution in \citetalias{2026ApJS..282...39J}.
At the characteristic scales investigated here ($R_{\text{c}} \sim 50-80$\,pc), CO molecules require substantial dust shielding against interstellar photodissociation. Consequently, dust continuum emission delineates a more extended envelope, incorporating both CO-dark molecular gas and atomic gas, which systematically yields larger cloud radii compared to CO-based boundaries, potentially overestimating the extent of the molecular cloud \citep[e.g.,][]{2026SCPMA..6969513X}.
For the cloud mass, \citetalias{2026ApJS..282...39J} derived the dust surface density by fitting a single-component modified blackbody spectral energy distribution (SED) to the \textit{Herschel} (70--500\,\textmu m) and JCMT SCUBA-2 (450 and 850\,\textmu m) observations. The total gas mass was then estimated by integrating the dust surface density over the physical area of the cloud, and applying a galactocentric radius-dependent dust-to-gas ratio, $g(R)$ \citep{Draine2014ApJ}. 
Similar to the radius, the GMC mass is likely overestimated because the dust continuum traces the total gas, which includes both molecular and atomic components.

With the cloud size and mass, we calculate the virial parameter, $\alpha_{\rm vir}$, as a measure of the dynamical state of each cloud. 
The virial parameter is computed following \citet{2021ApJ...920..126E},
\begin{equation}
\alpha_{\rm vir}
= \frac{2E_{\rm kin}}{|E_{\rm g}|}
= 1.16 \times 10^{3}
\left(\frac{\sigma_v}{\rm km\,s^{-1}}\right)^2
\left(\frac{R_{\rm c}}{\rm pc}\right)
\left(\frac{M_{\rm c}}{M_\odot}\right)^{-1}.
\label{equ:virpara}
\end{equation}
where $\sigma_{\rm v}$ is the channel-width-corrected velocity dispersion derived from the Gaussian fit to the cloud-averaged CO(1-0) spectrum, $R_{\rm c}$ is the cloud effective radius, and $M_{\rm c}$ is the cloud mass.
The uncertainty in $\alpha_{\rm vir}$ is propagated from those in $\sigma_v$, $R_{\rm c}$, and $M_{\rm c}$.
We adopt a 20\% uncertainty for the dust column density from \citetalias{2026ApJS..282...39J}.
For $R_{\rm c}$, we adopt a conservative 20\% uncertainty to account for uncertainties associated with the dendrogram-defined cloud boundary and finite-resolution effects, motivated by previous studies of cloud-property measurements \citep[e.g.,][]{2006PASP..118..590R}.
The uncertainty in $\sigma_v$ is taken from the Gaussian fit, with a median relative uncertainty of $\sim4\%$ for the final sample.
These terms result in a median propagated uncertainty of $\sim50\%$ in the $\alpha_{\rm vir}$.
These propagated uncertainties do not fully capture possible systematic overestimates in dust-defined $R_{\rm c}$ and $M_{\rm c}$, since dust traces the total gas, including both HI and H$_2$.

\begin{figure*}[!htb]
\centering
\begin{minipage}{0.32\linewidth}
    \centering
    \includegraphics[width=\textwidth]{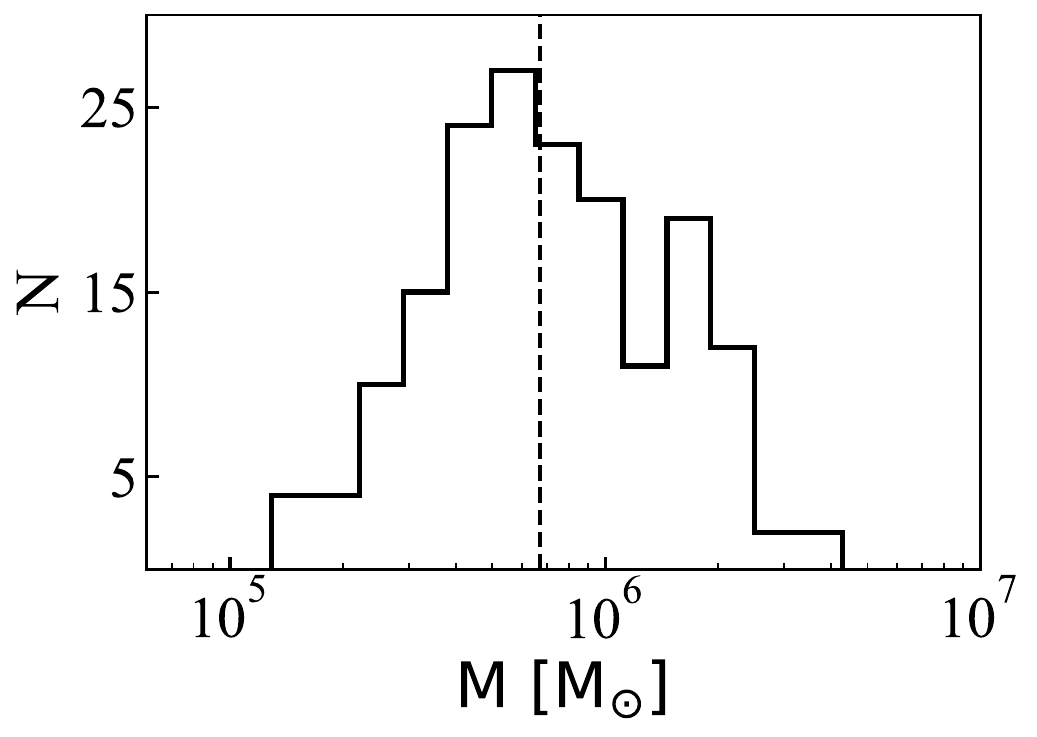}
\end{minipage}
\hfill
\begin{minipage}{0.32\linewidth}
    \centering
    \includegraphics[width=\textwidth]{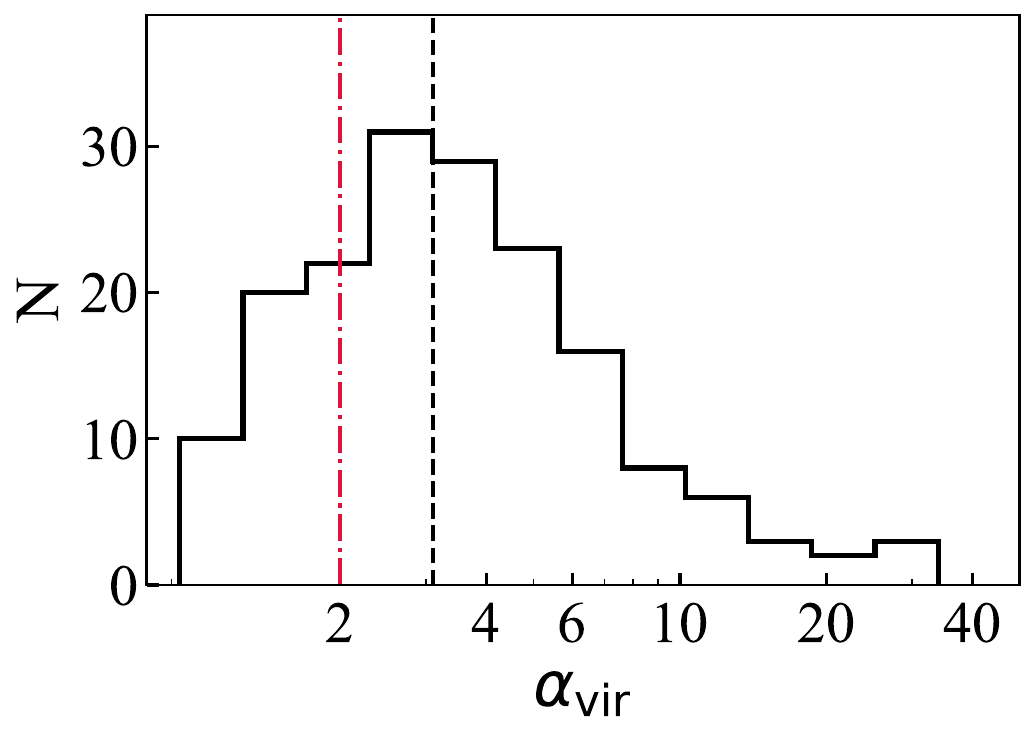}
\end{minipage}
\hfill
\begin{minipage}{0.32\linewidth}
    \centering
    \includegraphics[width=\textwidth]{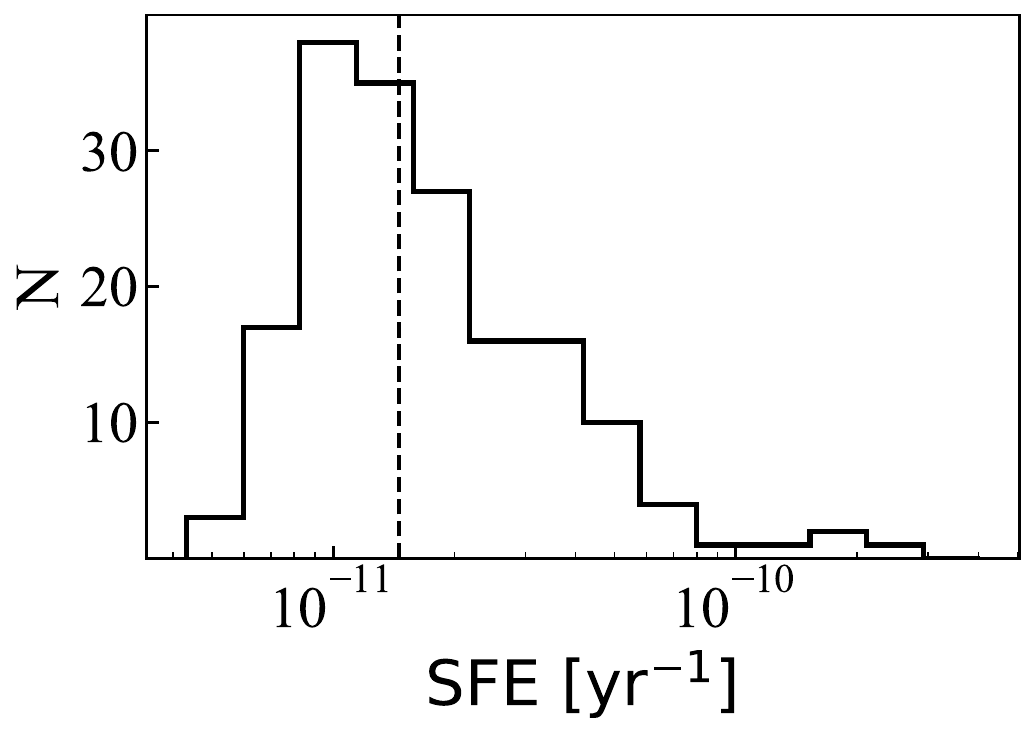}
\end{minipage}
\caption{Distributions of GMC masses (left panel), virial parameters (middle panel), and SFEs (right panel).
The black vertical dashed lines indicate the median values of the corresponding distributions, while the red dashed line in the middle panel denotes the threshold of $\alpha_{\rm vir}=2$.}
\label{fig:hist}
\end{figure*}

Furthermore, the SFE of each cloud is defined as the ratio between the SFR and the cloud mass:
\begin{equation}
\mathrm{SFE} = \frac{\mathrm{SFR}{\rm_c}}{M{\rm_c}},
\end{equation}
where $\mathrm{SFR}_{\rm c}$ is obtained by integrating the SFR surface density map of \citet{ford2013ApJ} over the GMC aperture, and $M_{\rm c}$ is the cloud mass.
Estimating the SFR at the scale of individual GMCs inherently introduces substantial uncertainties. 
Standard global SFR calibrations assume a fully sampled Initial Mass Function (IMF) and a continuous star formation history. 
At tens-of-parsec scales, however, stochastic sampling of the IMF can cause significant over- or underestimation of the instantaneous SFR. 
Furthermore, temporal mismatches between the visibility timescales of recent star formation tracers and typical GMC lifetimes introduce a large, two-way stochastic scatter into the derived SFE, a physical effect we will further discuss in Section \ref{subsec:sfe_vir}.
Uncertainties for all derived quantities are estimated using standard error propagation.
The derived quantities are all listed in Table \ref{tab:sample_data}.

With these physical properties determined, we first present their distributions in Figure \ref{fig:hist}, with median values indicated by black vertical dashed lines. 
As shown in the figure, the GMC masses predominantly range from $10^5$ to a few $10^6\,M_{\odot}$. 
The virial parameter distribution indicates that the majority of the clouds in our sample are gravitationally unbound, with $\alpha_{\text{vir}}$ values predominantly exceeding the critical threshold of $\sim 2$ (75.7\%). 
The derived SFE values span roughly three orders of magnitude, typically ranging from $\sim 10^{-12}$ to $10^{-9}$\,yr$^{-1}$, corresponding to gas depletion times of $\sim1-10^3$\,Gyr. 
For reference, molecular gas depletion times in nearby star-forming galaxies from PHANGS-ALMA are typically of order $\sim1-3$\,Gyr, with the broader distribution extending to $\sim0.1-10$\,Gyr \citep{2025ApJ...985...14L}. 
Thus, most GMCs in our M31 sample lie toward the low-SFE, long-depletion-time regime. 

\begin{figure*}
    \centering
        \includegraphics[width=0.8\linewidth]{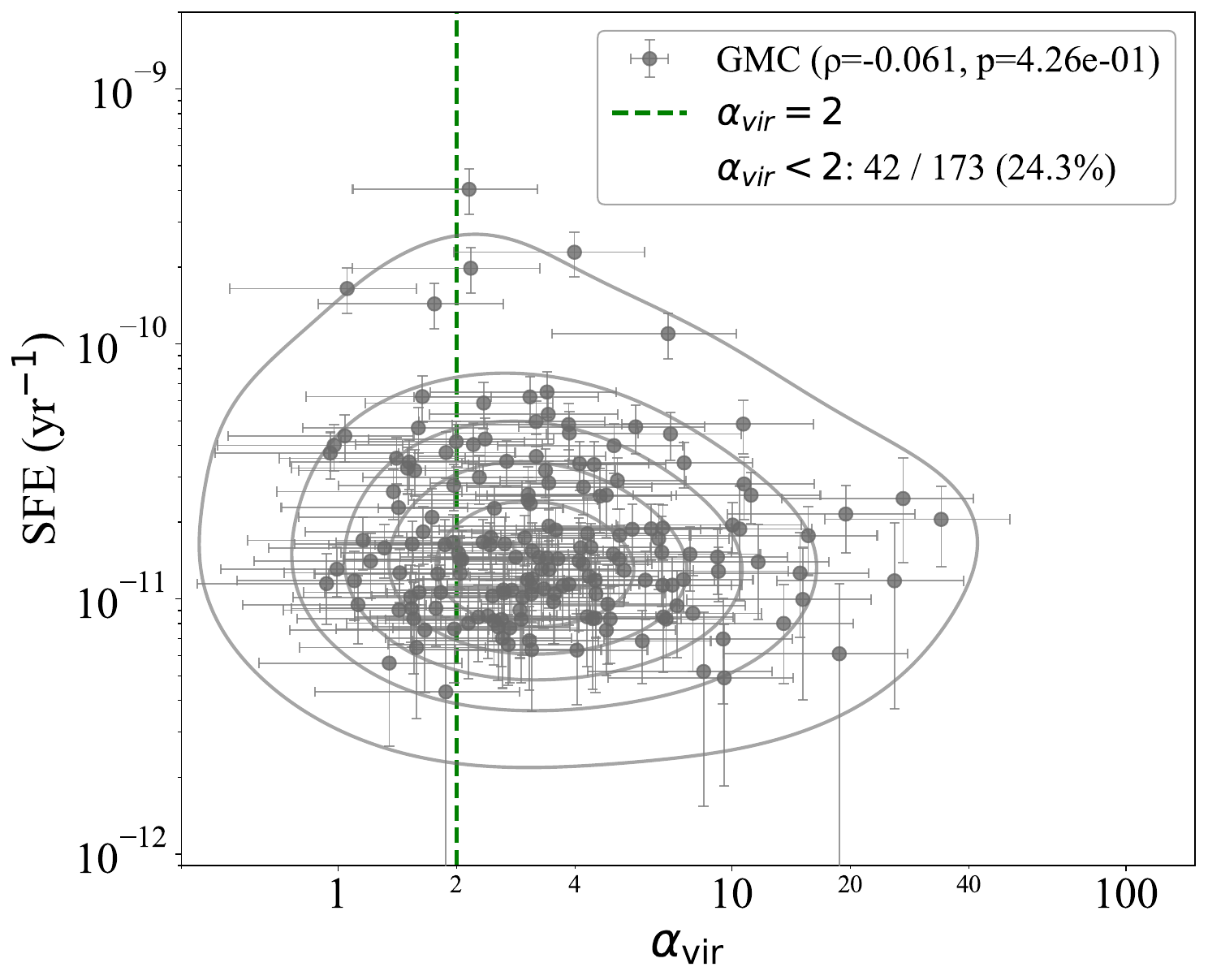}
        \caption{Relation between the SFE and the $\alpha_{\text{vir}}$. 
        The contours outline the distribution of the data points, decreasing in number density from the peak (center) to the 95\% probability boundary (outermost). 
        The vertical green dashed line marks the $\alpha_{\text{vir}} = 2$ threshold. The Spearman correlation coefficient ($\rho$) and the corresponding $p$-value are annotated in the legend. The error bars represent the uncertainties propagated from the measurements of mass, size, SFR, and $\sigma_v$.}
    \label{fig:sfe_vir}
\end{figure*}

Then, we examine the relationship between SFE and the virial parameter in Figure \ref{fig:sfe_vir}.
To quantify any potential correlation, we perform a Spearman rank-order correlation analysis on the sample. 
It yields a correlation coefficient of $\rho = -0.061$ with a $p$-value of $0.426$, indicating no statistically significant correlation.
We further assess the robustness of this result using a bootstrap resampling analysis.
In this test, we randomly resample the GMCs with replacement $10000$ times, recompute the Spearman coefficient for each resampled data set, and use the resulting distribution of $\rho$ to estimate the confidence interval.
The bootstrap analysis yields a 95\% confidence interval for the Spearman correlation coefficient of $[-0.211, 0.092]$. 
Since the confidence interval encompasses zero and the p-value exceeds the standard significance threshold of 0.05, we cannot reject the null hypothesis that there is no significant dependence of the SFE on the virial parameter for the GMCs in our sample.

\section{Discussion}\label{section:discuss}

\subsection{Virial Parameters}\label{subsec:boundstate}

\subsubsection{Numerical Distribution and Caveats}\label{subsubsec:vir_num}

As shown in Section~\ref{section:results}, a striking feature of our sample is the prevalence of high virial parameters.
The distribution shown in Figure~\ref{fig:hist} (middle panel) indicates that the majority ($\sim$76\%) of the identified GMCs exhibit $\alpha_{\rm vir} > 2$, suggesting that molecular clouds in M31 are largely gravitationally unbound at the $\sim86$-pc resolution probed by the CO(1-0) data.
This result is consistent with recent higher-resolution studies.
\citet{2024ApJ...966..193L}, using SMA CO(2-1) observations of M31 GMCs, found that only $\sim43\%$ of clouds are gravitationally bound, with the fraction decreasing to $\sim25\%$ for clouds lacking strong $^{13}$CO emission.
Similarly, the HASHTAG survey \citep{2025MNRAS.538.2445D}, based on JCMT CO(3-2) observations, reported mean virial parameters of $\alpha_{\rm vir} \sim 3.6$, despite CO(3-2) tracing relatively denser gas.
In the circum-nuclear region of M31, \citet{2019A&A...625A.148D} identified 12 CO(1-0) clumps with IRAM-PdBI observations at $\sim11$\,pc resolution and found a mean virial parameter of $\alpha_{\rm vir}=140\pm72$, suggesting that these central molecular structures are also not virialized.
These results indicate that elevated virial parameters are common for CO-bright molecular structures in M31 across different tracers, environments, and spatial scales.

However, the interpretation of these large $\alpha_{\rm vir}$ values requires caution.
The virial parameter assumes that clouds are isolated, approximately spherical systems with simple density structures, whereas real GMCs exhibit complex, hierarchical morphologies \citep{1993prpl.conf..125B,2023ASPC..534....1C}.
Moreover, CO emission, especially for low-J $^{12}$CO transitions, primarily traces the extended, diffuse envelopes of molecular clouds, which may be dynamically decoupled from the denser, collapsing regions.
Even so, CO linewidths remain a useful probe of cloud-scale kinematics in extragalactic studies.
For example, \citet{2008ApJ...686..948B} showed that GMCs identified via CO emission exhibit remarkably uniform properties across a wide range of galactic environments, suggesting that CO linewidths are primarily governed by macroscopic virialized motions.
In addition, \citet{2013ARA&A..51..207B} suggested that on GMC scales, the molecular clouds can be considered as ensembles of numerous discrete, clumpy structures which are sufficiently sparse in spatial-spectral space to avoid significant shadowing, making the ensemble behave as optically thin entity.

Nevertheless, the present virial analysis neglects complexities such as external pressure confinement \citep{2011MNRAS.416..710F,2025ApJ...986...12L} or the possibility that several structures are transient features in the process of dispersing \citep{2019A&A...625A.148D}. 
Given these caveats, $\alpha_{\text{vir}}$ derived from CO kinematics may systematically overestimate the true dynamical state of the gas, making it a potentially suboptimal indicator of the bound gas fraction involved in star formation.

\subsubsection{Spatial Distribution}\label{subsubsec:vir_spa}

\begin{figure}
    \includegraphics[width=\linewidth]{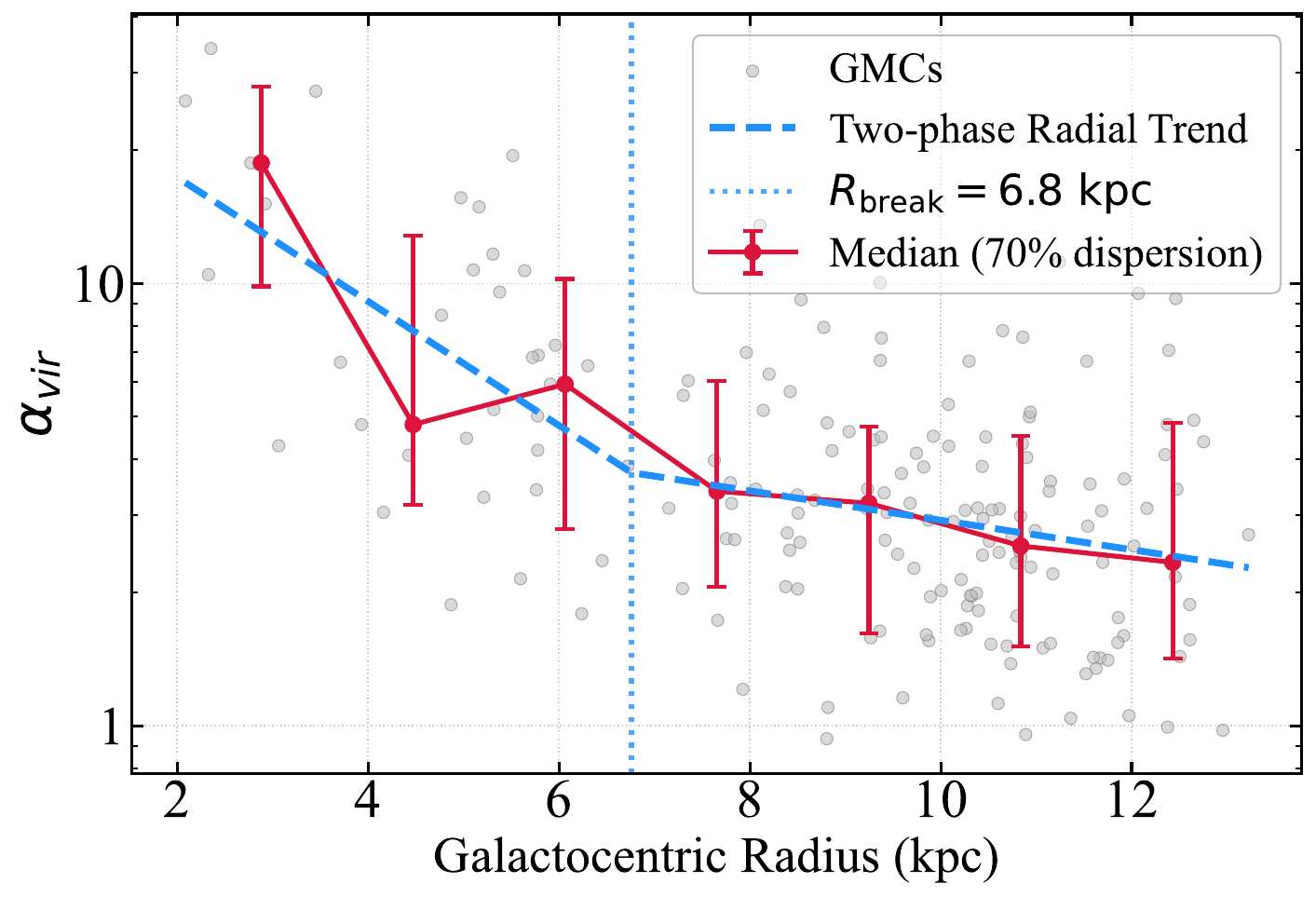}
    \caption{Radial variation of the GMC virial parameter as a function of galactocentric radius in M31. Gray circles represent individual GMCs in our sample. The red symbols and associated error bars denote the median values and the 70\% dispersion (corresponding to the 15th--85th percentiles) calculated within sliding radial bins. 
    The blue dashed line illustrates a two-phase linear fit to the logarithmic data, with an apparent transition near $R \sim 6$-7 kpc marked by the vertical blue dotted line.
    }
    \label{fig:vir_galacR}
\end{figure}

\begin{figure}
    \includegraphics[width=\linewidth]{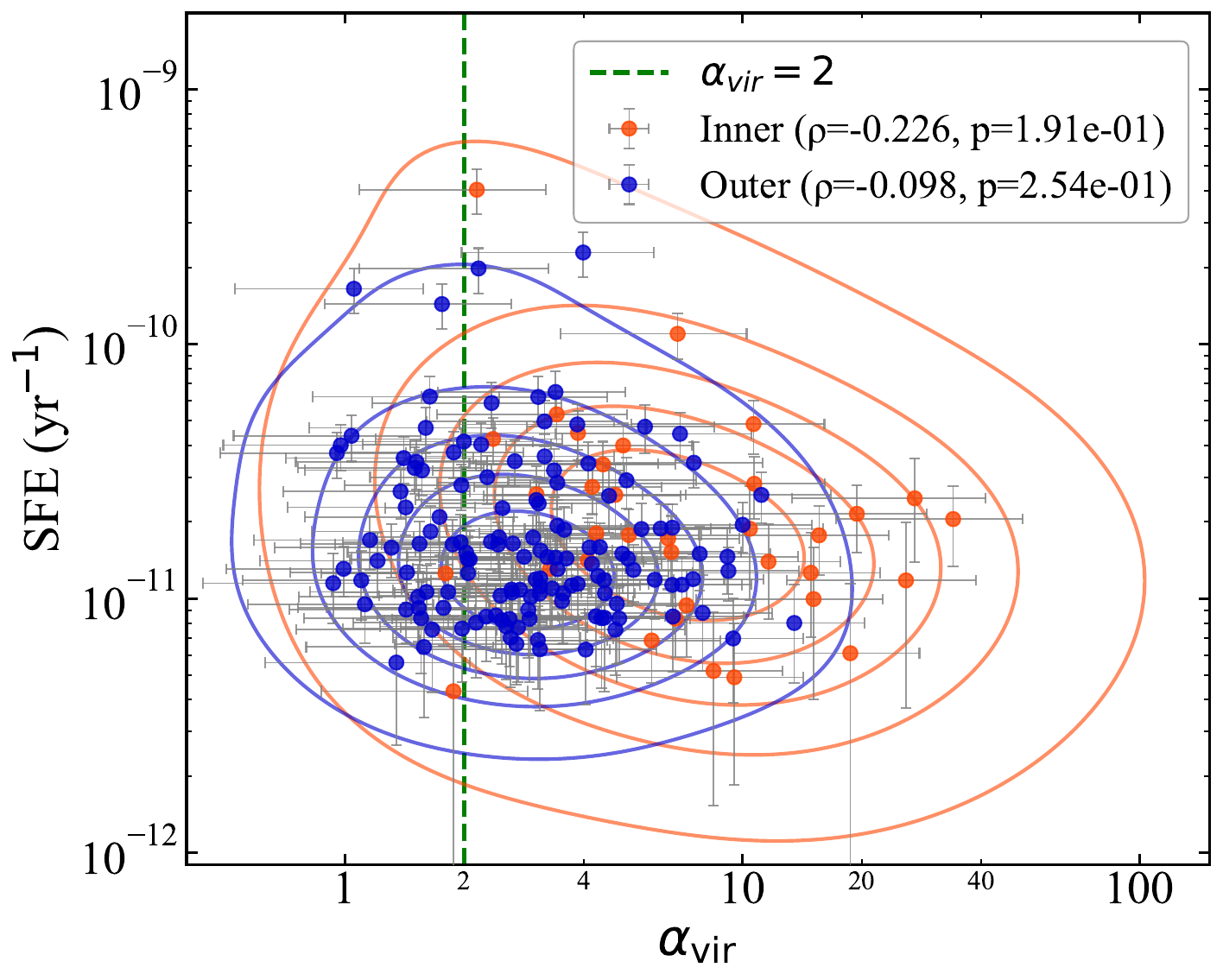}
    \caption{Relation between the SFE and the $\alpha_{\text{vir}}$ for GMC subsamples. The clouds are classified into two populations based on their galactocentric locations: the Inner Region ($R \le 6.8\,kpc$, orange) and the Outer Region ($R > 6.8\,kpc$, mediumblue). For each population, the Spearman rank-order correlation coefficient $\rho$ and the corresponding p-value are annotated in the legend. The overlaid contours represent the distribution of each subgroup, decreasing in number density from the peak (center) to the 95\% probability boundary (outermost). The vertical dashed line marks the $\alpha_{\text{vir}} = 2$ threshold.}
\label{fig:virsfe_subgroup}
\end{figure}

Beyond the overall distribution, we find systematic spatial variations in $\alpha_{\rm vir}$ across the M31 disk. 
As shown in Figure~\ref{fig:vir_galacR}, the virial parameter shows an apparent radial variation: it decreases from the inner disk to $R \sim 6$-7\,kpc and becomes flatter at larger radii.
This behavior is also reflected in the SFE–$\alpha_{\rm vir}$ plane (Figure~\ref{fig:virsfe_subgroup}), where GMCs in the inner regions systematically occupy higher $\alpha_{\rm vir}$ values than those in the outer disk.

This radial variation likely reflects the influence of the large-scale galactic environment.
The transition at $R \sim 6$–7\,kpc approximately corresponds to the boundary between the bulge-dominated inner region and the main star-forming disk, which includes the prominent ring at $R \approx 5.6$\,kpc \citep{Draine2014ApJ}.
In the inner few kiloparsecs, the deep gravitational potential of the central bulge and the steeply rising rotation curve induce strong galactic shear \citep{2009ApJ...705.1395C}.
According to the model of \citet{2018ApJ...854..100M}, motions driven by such galactic gravitational potentials can resemble or even exceed the motions required for self-gravity support. 
In addition, differential rotation in the disk \citep{2020MNRAS.496.5211A} can inject kinetic energy into GMCs, increasing their internal velocity dispersions and, consequently, their virial parameters.

Other radial variations in the ISM conditions of M31 point in the same direction.
The CO(3-2)/CO(1-0) line ratio across the M31 disk declines from the nuclear region to intermediate radii and rises again in the 10\,kpc ring \citep{2020MNRAS.492..195L}.
In the circumnuclear region, the molecular gas is also found to have a higher temperature than that in the disk \citep{2019MNRAS.484..964L}, while far-infrared fine-structure line observations show a strong radiation field in the inner galaxy \citep{2020ApJ...905..138L}.
These results indicate that the physical conditions of the molecular ISM vary systematically across the disk, consistent with the radial variation seen in $\alpha_{\rm vir}$.

A related effect may arise from the metallicity and environmental gradients across the M31 disk.
\citet{2025NatAs...9..406L} reported a similar trend in the Milky Way, where molecular clouds in the metal-rich inner Galaxy tend to exhibit higher virial parameters than those in the outer Galaxy, even after accounting for metallicity-dependent mass estimates.
The elevated $\alpha_{\rm vir}$ values in the inner disk of M31 may therefore reflect a more general environmental dependence of cloud dynamical states in metal-rich, dynamically active regions.
In this context, the radial variation of $\alpha_{\rm vir}$ in M31 is likely not driven by metallicity alone, but instead reflects the combined influence of metallicity, large-scale dynamics, and local ISM conditions across the disk.

\subsubsection{Comparison with Star-Forming Galaxies}\label{subsec:comparePHANGS}

To assess whether GMCs in M31 differ systematically from those in normal star-forming galaxies, we compare their dynamical states with those in the PHANGS-ALMA survey \citep{2021ApJS..257...43L}.

\begin{figure}
    \includegraphics[width=\linewidth]{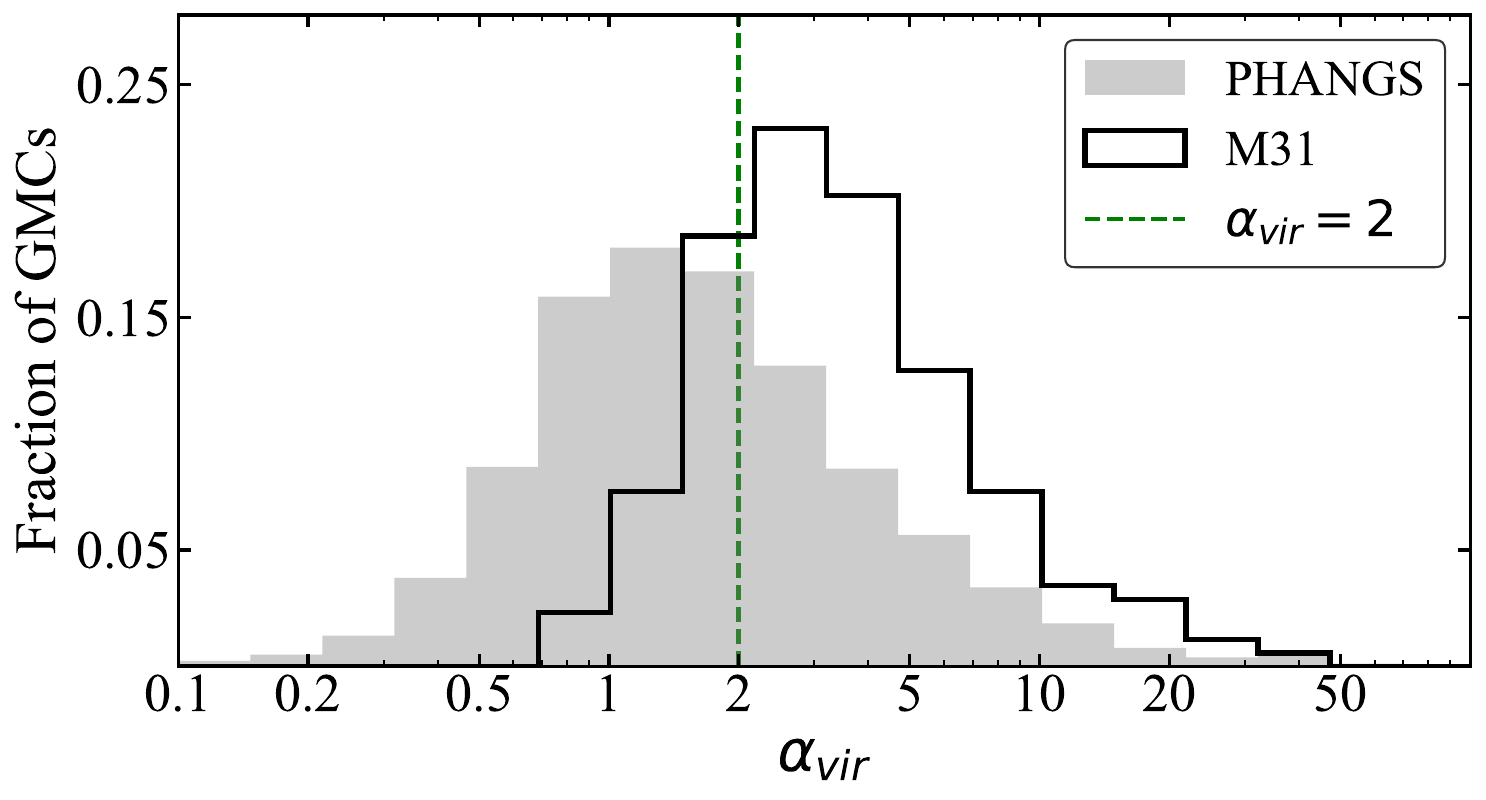}
    \caption{Normalized distributions of $\alpha_{\text{vir}}$ for the GMCs in M31 (black stepped line; this work) compared to those in 10 nearby main-sequence star-forming galaxies from the PHANGS-ALMA survey (gray shaded area; \citealt{2021MNRAS.502.1218R}). The vertical green dashed line denotes $\alpha_{\text{vir}} = 2$.}
\label{fig:phangsHist}
\end{figure}

We adopt the GMC catalog from \citet{2021MNRAS.502.1218R}, which identified individual clouds across 10 PHANGS-ALMA targets using the \textsc{pycprops} algorithm \citep{2006PASP..118..590R} after homogenizing the CO(2-1) data to a common physical resolution of 90\,pc.
Following their methodology, we evaluate the virial parameters of these GMCs as $\alpha_{\text{vir}} = 10 \sigma_v^2 R_{\text{3D}} / (G M_{\text{CO}})$, where $\sigma_v$ is the velocity dispersion, $R_{\text{3D}}$ is their estimated 3D cloud radius, and $M_{\text{CO}}$ is the CO-based luminous mass.
As shown in Figure~\ref{fig:phangsHist}, the PHANGS GMCs exhibit a distribution that peaks at $\alpha_{\rm vir} \sim 1$–2, whereas the M31 sample is systematically shifted toward higher values. 
This offset suggests that self-gravity plays a less dominant role in M31 clouds compared to those in actively star-forming galaxies.

We also note that the M31 CO(1-0) peak temperatures, when scaled to the CO(2-1) transition using a representative line ratio of $R_{21}=0.65$ \citep{Nieten2006A&A}, are systematically lower than the PHANGS CO(2-1) peak temperatures measured at similar physical resolution \citep{2018ApJ...860..172S}.
Low peak temperatures measured at tens-of-parsec scales are often interpreted as signatures of low beam-filling factors, indicating unresolved clumpy substructure within the telescope beam \citep[e.g.,][]{2013AJ....146...19L,2016ApJ...831...16L}.
In this context, the low $T_{\rm peak}$ values of M31 clouds may suggest a lower CO beam-filling factor, which could reflect a more fragmented or diffuse molecular gas distribution on sub-resolution scales.
Such a scenario is qualitatively consistent with expectations from turbulent ISM models, where stronger turbulence leads to broader density distributions and a reduced volume-filling fraction of dense gas \citep[e.g.,][]{2004RvMP...76..125M,2010A&A...512A..81F}.
If so, the elevated cloud-averaged virial parameters in M31 could partly reflect a reduced filling fraction of dense or self-gravitating gas within the CO beam.

\subsection{The Relationship between SFE and GMC Dynamics}\label{subsec:sfe_vir}

\subsubsection{The Global Trend}\label{subsubsec:sfe_vir_all}

We investigate the relationship between the star formation efficiency and the virial parameter for the GMC population in our sample.
As shown in Figure~\ref{fig:sfe_vir}, we do not find a statistically significant global correlation between SFE and $\alpha_{\rm vir}$.
This result is based on the Spearman rank-order coefficient and is further supported by bootstrap resampling, for which the 95\% confidence interval includes zero.
A correlation might be expected if star formation efficiency were directly regulated by the global dynamical state of molecular clouds.
However, the absence of such a monotonic trend in our data suggests that the connection between cloud-scale dynamics and SFE at tens-of-parsec scales is likely more complex.

A negative correlation may be expected under a framework in which star formation is regulated by the fraction of gravitationally bound gas within molecular clouds \citep[e.g.,][]{2005ApJ...630..250K,2021ApJ...920..126E}.
Observational studies of Milky Way clouds support this picture, showing that star formation is closely linked to the mass of self-gravitating gas \citep{2025A&A...701A.152J, 2025ApJ...987...77J}.
In this context, the globally low SFE observed in M31 may reflect a deficit of gravitationally bound gas within its GMCs.
If a large fraction of the gas remains dynamically unbound, the clouds would be characterized by elevated virial parameters and a reduced ability to collapse and form stars.
Under such conditions, an anti-correlation between SFE and $\alpha_{\rm vir}$ would naturally be predicted.
However, our results indicate that such a trend is not evident in our GMC sample.
At larger spatial scales, PHANGS measurements averaged over 1.5-kpc regions across nearby galaxies likewise show no clear dependence of the star formation efficiency per free-fall time, $\epsilon_{\rm ff}$, on the population-averaged cloud-scale virial parameter \citep{2024ARA&A..62..369S}.
Although $\epsilon_{\rm ff}$ is not identical to the SFE considered here, this result similarly indicates that the observed dynamical state of molecular gas does not translate into a simple monotonic variation in star formation efficiency.

One likely explanation lies in the limitations of the virial parameter as a proxy for the bound gas fraction.
As discussed in Section \ref{subsubsec:vir_num}, $\alpha_{\rm vir}$ represents a single, cloud-averaged quantity and does not capture the internal density and kinematic structure where star formation actually occurs.
Moreover, $\alpha_{\rm vir}$ does not uniquely distinguish between dynamical equilibrium and gravitational collapse.
Under idealized assumptions, $\alpha_{\rm vir}=1$ is consistent with a globally non-collapsing state of virial equilibrium and therefore does not, by itself, imply active star formation.
By contrast, dynamically collapsing clouds can exhibit $\alpha_{\rm vir}\sim2$ when collapse-driven motions, including infall, contribute substantially to the measured kinetic energy \citep[e.g.,][]{2022MNRAS.515.2822R}.
Accordingly, $\alpha_{\rm vir}<2$ does not identify clouds that must be actively collapsing, while $\alpha_{\rm vir}\sim2$ does not exclude ongoing collapse.
As a result, the inability of the cloud-averaged $\alpha_{\rm vir}$ to capture the internal density and kinematic structure of GMCs, combined with its dynamical degeneracy, precludes a unique correspondence with instantaneous SFE.

Another important factor is the difficulty of defining a meaningful SFE on the scale of individual clouds. 
Star formation is inherently time-dependent, and molecular gas tracers and recent star formation tracers probe different evolutionary phase of a cloud \citep[e.g.,][]{McKee2007ARA&A, 2020MNRAS.493.2872C}. 
This temporal mismatch between gas and young stellar population tracers introduces a significant, stochastic scatter in the estimated SFE, which can be averaged down over large spatial scales encompassing many clouds in different evolutionary stages \citep{2014MNRAS.439.3239K, Sun2023ApJ...945L..19S}. 

Taken together, these effects suggest that the intrinsic scatter in SFE and the limitations of $\alpha_{\rm vir}$ as a diagnostic can obscure any simple relationship between cloud dynamics and star formation efficiency at the scale of individual GMCs.
However, the absence of a global correlation does not preclude the possibility that $\alpha_{\rm vir}$ still constrains the range of SFE values accessible to a cloud population.

\subsubsection{The Upper-Tail Trend}\label{subsubsec:sfe_vir_up}

Although no statistically significant global correlation is found, the distribution of points in Figure~\ref{fig:sfe_vir} is not entirely random.
The highest-SFE clouds appear to occur preferentially at relatively low $\alpha_{\rm vir}$, suggesting that cloud dynamics may still affect the upper range of SFE values accessible to the GMC population.
To quantify this behavior, we perform quantile regression in the $\log{\rm SFE}$--$\log\alpha_{\rm vir}$ plane \citep{Koenker1978}.
For a given quantile level $\tau$, the fitted relation is
\begin{equation}
    Q_{\tau}(\log{\rm SFE}\,|\,\log\alpha_{\rm vir}) = a_{\tau}\log\alpha_{\rm vir} + b_{\tau},
\end{equation}
where $Q_{\tau}(\log{\rm SFE}\,|\,\log\alpha_{\rm vir})$ denotes the $\tau$th conditional quantile of the $\log{\rm SFE}$ distribution at fixed $\log\alpha_{\rm vir}$.
The parameters $a_{\tau}$ and $b_{\tau}$ are obtained by minimizing the standard asymmetric absolute-deviation loss for quantile regression.
All fits reported below converged successfully.

\begin{figure}[t!]
    \includegraphics[width=\linewidth]{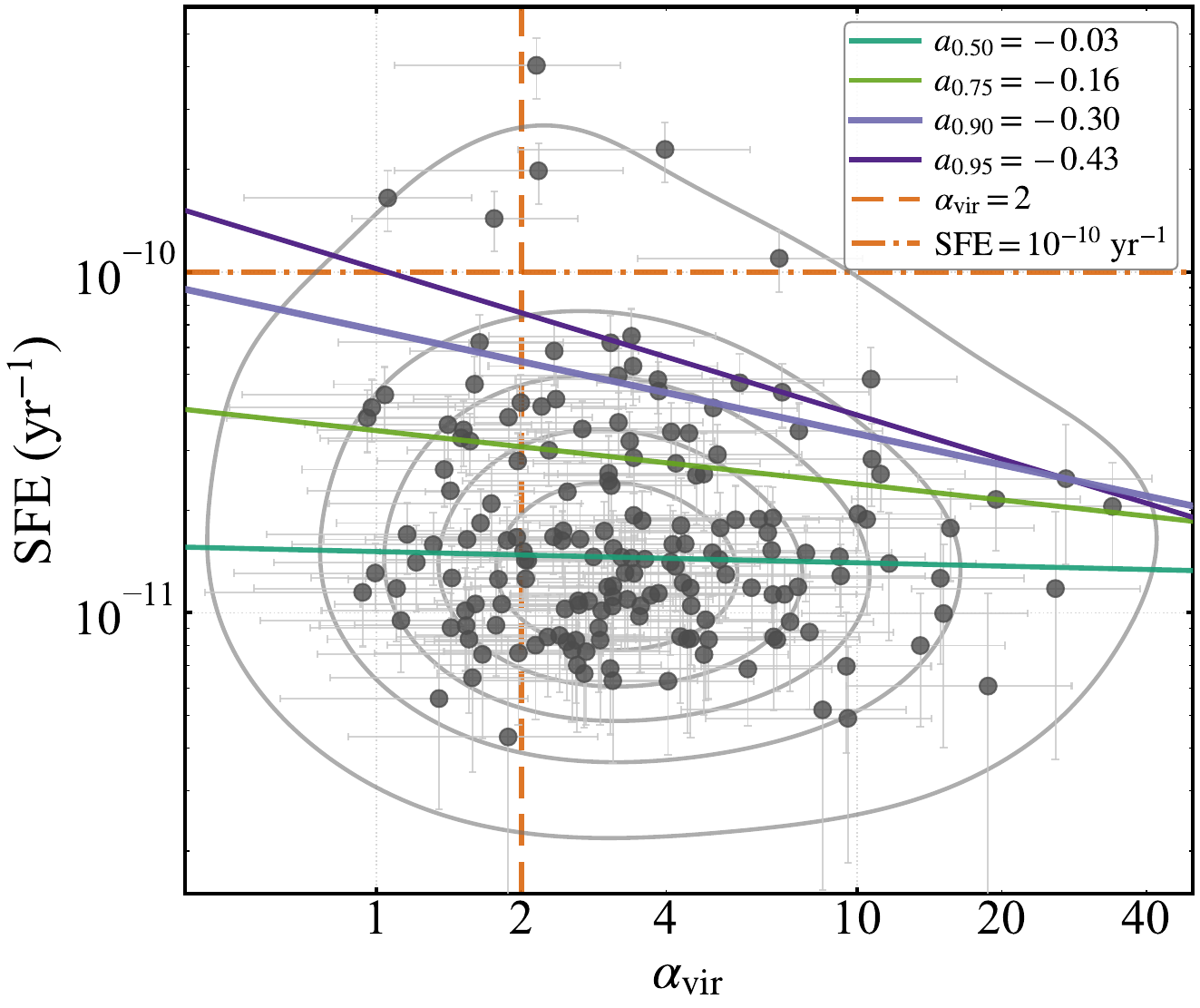}
    \caption{Relation between SFE and $\alpha_{\rm vir}$ with quantile-regression fits. 
    The data points and contours are the same as in Figure~\ref{fig:sfe_vir}. 
    Colored solid lines show linear quantile-regression fits in the $\log{\rm SFE}$--$\log\alpha_{\rm vir}$ plane for $\tau=0.50$, 0.75, 0.90, and 0.95, where $\tau$ denotes the conditional quantile of the SFE distribution at fixed $\alpha_{\rm vir}$. 
    The vertical orange dashed line marks the $\alpha_{\rm vir}=2$ threshold, while the horizontal orange dash-dotted line indicates $SFE=10^{-10}$~yr$^{-1}$.
    }
\label{fig:sfe_vir_up}
\end{figure}

Figure~\ref{fig:sfe_vir_up} shows the fitted quantile-regression relations for $\tau=0.50$, 0.75, 0.90, and 0.95.
The median relation is nearly flat, with $a_{0.50}=-0.03$.
The fitted slopes become progressively more negative toward higher quantiles, with $a_{0.75}=-0.16$, $a_{0.90}=-0.30$, and $a_{0.95}=-0.43$.
The contrast between the nearly flat median relation and the steeper upper-quantile slopes indicates that the SFE-$\alpha_{\rm vir}$ dependence, if present, is not a population-wide monotonic trend but is instead confined mainly to the high-SFE end of the distribution.

We use the $\tau=0.90$ fit as a fiducial upper-tail statistic, because it probes the high-SFE part of the distribution while remaining less sensitive to the most extreme points than the $\tau=0.95$ fit.
To test whether the observed upper-tail slope could arise from the skewed marginal distributions alone, we performed a permutation test in which the SFE values were randomly shuffled relative to $\alpha_{\rm vir}$ while preserving both marginal distributions.
Only 1.2\% of the randomized realizations produce a $\tau=0.90$ slope as negative as or more negative than the observed value.
This indicates that the declining upper tail is unlikely to be produced solely by the skewed marginal distributions of SFE and $\alpha_{\rm vir}$.

The declining upper-tail trend may arise from two effects.
First, the cloud-averaged virial parameter may influence the maximum SFE values reached by the GMC population.
Clouds with large $\alpha_{\rm vir}$ are expected to contain a smaller fraction of self-gravitating gas on average, reducing their ability to maintain dense, collapsing substructures.
Conversely, clouds with lower $\alpha_{\rm vir}$ may be more capable of sustaining self-gravitating substructure and can therefore reach higher instantaneous SFE values.
This interpretation is qualitatively consistent with turbulence-regulated star formation models and numerical simulations \citep[e.g.,][]{2012ApJ...759L..27P,2015MNRAS.450.4035F}, which predict reduced star formation activity in dynamically unbound gas.

Second, the observed CO-selected sample may underrepresent GMCs in the feedback-dispersal phase.
As stars form, stellar feedback can broaden, disrupt, heat, or dissociate the surrounding molecular gas.
Once feedback has substantially affected the natal gas, the dispersed material may no longer be CO-bright, and the remaining molecular emission may no longer be identified as a coherent GMC in our selection.
High-SFE clouds whose molecular gas has already been strongly disrupted would therefore be difficult to include in a CO-selected virial analysis.
This selection effect can suppress the number of high-SFE, high-$\alpha_{\rm vir}$ objects and may contribute to the declining upper tail seen in Figure~\ref{fig:sfe_vir_up}.

Below the upper-tail trend, the data exhibit substantial scatter over the full range of $\alpha_{\rm vir}$.
In addition to the scatter introduced by the temporal mismatch between gas and star formation tracers, variations associated with different evolutionary stages of GMCs are also likely to contribute to the observed dispersion.
During a cloud lifecycle, the instantaneous SFE can vary substantially as molecular gas is assembled, collapses to form stars, and is subsequently dispersed by feedback \citep[e.g.,][]{2009ApJS..184....1K,2014MNRAS.439.3239K,2020MNRAS.493.2872C,Meng2026}.
The virial parameter may also evolve during this sequence, as the balance among self-gravity, turbulence, magnetic fields, and external pressure changes with time \citep[e.g.,][]{2016ApJ...833..229L,2017ApJ...845..133S}.
However, the evolution of SFE and $\alpha_{\rm vir}$ may not be synchronized for individual clouds, since they are inferred from different tracers and respond to different phases and timescales of cloud evolution.
A further complication is that $\alpha_{\rm vir}$ is also affected by the larger-scale galactic environment.
As shown in Section~\ref{subsubsec:vir_spa}, GMCs in the inner disk of M31 have systematically elevated $\alpha_{\rm vir}$ values, indicating that environmental conditions can also influence the cloud-averaged dynamical state.
As a result, clouds with similar cloud-averaged $\alpha_{\rm vir}$ can have different recent star formation histories, while clouds with similar SFE can occupy different dynamical states.

The small number of high-SFE clouds in our sample is consistent with the view that low $\alpha_{\rm vir}$ is not, by itself, a sufficient condition for high instantaneous SFE.
Adopting a threshold of $10^{-10}\ {\rm yr}^{-1}$, corresponding to a molecular gas depletion time of 10\,Gyr, we find that only 3.5\% of the sample exceeds this value.
Although this fraction cannot be directly converted into an evolutionary timescale, it is broadly consistent with the picture that vigorous star formation represents a relatively short-lived phase in the GMC lifecycle \citep[e.g.,][]{2009ApJS..184....1K,2020MNRAS.493.2872C}.

\begin{deluxetable*}{lcccccccccc}
\tablecaption{Physical Properties of Selected Sources\label{tab:sample_data}}
\tablehead{
\colhead{Index} & \colhead{On\_Arm} & \colhead{RA} & \colhead{DEC} & \colhead{R} & \colhead{R$_c$} & \colhead{$\Sigma_{\text{SFR}}$} & \colhead{M$_c$} & \colhead{$\sigma_v$} & \colhead{SFE} & \colhead{$\alpha_{\text{vir}}$} \\
\colhead{} & \colhead{} & \colhead{(J2000)} & \colhead{(J2000)} & \colhead{(kpc)} & \colhead{(pc)} & \colhead{(M$_\sun$ yr$^{-1}$ kpc$^{-2}$)} & \colhead{(10$^5$ M$_\sun$)} & \colhead{(km s$^{-1}$)} & \colhead{(yr$^{-1}$)} & \colhead{}
}
\startdata
J9   & True  & 00:41:20.59 & 40:40:43.0 & 12.389 & 74.398 & 0.0007 & 11.30 & 9.6 & $1.1 \times 10^{-11}$ & 7.1 \\
J10  & True  & 00:41:13.90 & 40:40:50.0 & 11.686 & 77.749 & 0.0005 & 13.89 & 6.9 & $6.9 \times 10^{-12}$ & 3.1 \\
J11  & True  & 00:40:31.33 & 40:41:27.0 & 9.361  & 97.676 & 0.0041 & 19.91 & 5.4 & $6.2 \times 10^{-11}$ & 1.6 \\
J13  & True  & 00:41:21.27 & 40:41:23.0 & 12.071 & 45.957 & 0.0004 & 3.62  & 8.0 & $7.0 \times 10^{-12}$ & 9.5 \\
J14  & True  & 00:40:38.36 & 40:41:32.0 & 9.378  & 48.276 & 0.0009 & 5.82  & 8.8 & $1.2 \times 10^{-11}$ & 7.5 \\
J20  & True  & 00:41:34.94 & 40:43:16.0 & 12.374 & 47.515 & 0.0006 & 5.23  & 6.7 & $7.5 \times 10^{-12}$ & 4.8 \\
J21  & True  & 00:41:38.79 & 40:43:60.0 & 12.351 & 71.401 & 0.0019 & 8.93  & 6.7 & $3.4 \times 10^{-11}$ & 4.1 \\
J30  & True  & 00:40:40.68 & 40:45:20.0 & 8.376  & 71.401 & 0.0007 & 7.92  & 4.4 & $1.4 \times 10^{-11}$ & 2.1 \\
J32  & True  & 00:40:59.33 & 40:45:25.0 & 8.497  & 47.515 & 0.0007 & 4.39  & 5.1 & $1.1 \times 10^{-11}$ & 3.3 \\
J35  & True  & 00:40:43.45 & 40:46:28.0 & 8.104  & 53.295 & 0.0004 & 4.51  & 9.9 & $8.0 \times 10^{-12}$ & 13.5\\
J36  & True  & 00:41:46.47 & 40:46:36.0 & 11.628 & 43.515 & 0.0003 & 3.26  & 3.0 & $5.6 \times 10^{-12}$ & 1.3 \\
J44  & True  & 00:41:3.48  & 40:47:38.0 & 7.796  & 53.295 & 0.0005 & 4.22  & 4.9 & $1.0 \times 10^{-11}$ & 3.5 \\
J46  & True  & 00:41:7.34  & 40:48:02.0 & 7.754  & 54.644 & 0.0008 & 4.74  & 4.5 & $1.6 \times 10^{-11}$ & 2.7 \\
J56  & True  & 00:41:21.73 & 40:49:47.0 & 7.648  & 66.653 & 0.0045 & 9.78  & 6.6 & $6.5 \times 10^{-11}$ & 3.4 \\
J59  & True  & 00:41:25.96 & 40:49:59.0 & 7.809  & 73.907 & 0.0038 & 13.23 & 7.0 & $5.0 \times 10^{-11}$ & 3.2 \\
J60  & True  & 00:40:42.60 & 40:50:16.0 & 7.842  & 64.993 & 0.0006 & 6.92  & 4.9 & $1.1 \times 10^{-11}$ & 2.6 \\
J63  & True  & 00:41:30.16 & 40:50:55.0 & 7.628  & 80.056 & 0.0079 & 6.95  & 5.5 & $2.3 \times 10^{-10}$ & 4.0 \\
J64  & True  & 00:40:41.51 & 40:51:04.0 & 7.964  & 54.644 & 0.0020 & 4.27  & 6.9 & $4.4 \times 10^{-11}$ & 7.0 \\
J66  & True  & 00:42:13.87 & 40:51:21.0 & 11.973 & 107.271& 0.0087 & 19.02 & 4.0 & $1.6 \times 10^{-10}$ & 1.1 \\
J74  & False & 00:40:35.10 & 40:52:43.0 & 8.818  & 105.215& 0.0007 & 19.68 & 4.2 & $1.2 \times 10^{-11}$ & 1.1 \\
\enddata
\tablecomments{This table presents a subset of the final sample. The full version is available in machine-readable form. Column definitions are as follows: 
(1) Index: Source ID compiled in \citetalias{2026ApJS..282...39J}. 
(2) On\_Arm: Whether the source is located on a spiral arm. 
(3) RA: Right Ascension (J2000). 
(4) DEC: Declination (J2000). 
(5) R: Galactocentric Distance. 
(6) R$_c$: GMC Radius. 
(7) $\Sigma_{\text{SFR}}$: Surface Density of Star Formation Rate. 
(8) M: GMC mass. 
(9) $\sigma_v$: Velocity dispersion. 
(10) SFE: Star Formation Efficiency. 
(11) $\alpha_{\text{vir}}$: Virial parameter.}
\end{deluxetable*}


\section{Summary and Conclusions}\label{section:conclusion}
In this work, we utilized the comprehensive GMC catalog identified from JCMT-SCUBA2 850\,\textmu m continuum emission \citep{2026ApJS..282...39J} and CO(1-0) observations \citep{Nieten2006A&A} to investigate the physical properties and dynamical states of 173 GMCs in M31. We derived the virial parameters and star formation efficiencies for the sample and analyzed their correlations. 
The main results are summarized as follows:

\begin{enumerate}

\item The majority of GMCs in our sample ($\sim$76\%) exhibit high virial parameters, with a median value of $\alpha_{\rm vir}=3.1$ and a mean value of $\alpha_{\rm vir}=4.6$. This suggests that the GMC population in M31 is largely gravitationally unbound at the $\sim$\,86\,pc resolution probed by the CO(1-0) data.
The virial parameter exhibits a radial gradient, being elevated in the dynamically active inner region ($R \lesssim 7$\,kpc) and flattening out in the outer disk.
Moreover, comparisons with the PHANGS-ALMA survey reveal that M31 GMCs possess systematically higher $\alpha_{\text{vir}}$ and lower $T_{\text{peak}}$ values, demonstrating that GMCs in M31 are less gravitationally bound than those in star-forming galaxies.

\item We find no statistically significant global correlation between SFE and $\alpha_{\rm vir}$ for the full GMC sample, indicating that the cloud-averaged virial parameter does not provide a one-to-one predictor of the instantaneous SFE of individual GMCs. 
A quantile-regression analysis further shows that the median SFE-$\alpha_{\rm vir}$ relation is nearly flat, while the upper quantiles decline toward larger $\alpha_{\rm vir}$. 
This behavior suggests that $\alpha_{\rm vir}$ may constrain the upper range of SFE values accessible to the GMC population, while the large scatter in the SFE-$\alpha_{\rm vir}$ plane reflects additional effects such as cloud evolution, tracer timescale mismatch, and unresolved internal cloud structure.

\end{enumerate}

Future high-resolution observations of M31 GMCs will be required to resolve their internal density and kinematic structures and to move beyond cloud-averaged virial parameters. 
Such data would enable more direct diagnostics of self-gravitating gas, including N-PDF analysis, which can isolate the high-column-density power-law tail associated with gravitational collapse, and differential virial analysis, which characterizes the dynamical state as a function of spatial scale. 
These measurements will provide a more direct test of whether the low SFE in M31 is linked to a deficit of gravitationally bound gas within GMCs.

\begin{acknowledgements}
This work is supported by NSFC grant nos. 12588202 and 12041302, by the National Key R\&D Program of China No. 2023YFA1608004.
D. L. acknowledges support from the New Cornerstone foundation.
\end{acknowledgements}


\bibliographystyle{aasjournal}
\bibliography{reference}

\end{document}